\documentstyle{article}
\begin{document}
{\Large The Case Against Cosmology}

\vspace{5mm}

M. J. Disney\footnote{Physics and Astronomy, Cardiff
    University, Cardiff CF24 3YB, Wales, UK}

\vspace{10mm}

\begin{abstract}
It is argued that some of the recent claims for cosmology are grossly
overblown.  Cosmology rests on a very small database: it suffers from
many fundamental difficulties as a science (if it is a science at all)
whilst observations of distant phenomena are difficult to make and
harder to interpret.  It is suggested that cosmological inferences
should be tentatively made and sceptically received
\end{abstract}

\section{INTRODUCTION}

Given statements emanating from some cosmologists today one could be
forgiven for assuming that the solution to some of the great problems
of the subject, even ``the origin of the Universe'' lie just around
the corner.  As an example of this triumphalist approach consider the
following conclusion from Hu et al. [1] to a preview of the results
they expect from spacecraft such as MAP and PLANCK designed to map the
Cosmic Background Radiations: ``\ldots we will establish the
cosmological model as securely as the Standard Model of elementary
particles.  We will then know as much, or even more, about the early
Universe and its contents as we do about the fundamental constituents
of matter''. 

We believe the most charitable thing that can be said of such
statements is that they are naive in the extreme and betray a complete
lack of understanding of history, of the huge difference between an
observational and an experimental science, and of the peculiar
limitations of cosmology as a scientific discipline.  By building up
expectations that cannot be realised, such statements do a disservice
not only to astronomy and to particle physics but they could
ultimately do harm to the wider respect in which the whole scientific
approach is held.  As such, they must not go unchallenged.

It is very questionable whether the study of any phenomenon that is not
repeatable can call itself a science at all.  It would be sad however
to abandon the whole fascinating area to the priesthood.  But if we
are going to lend this unique subject any kind of scientific
respectability we have to look at all its claims with a great
circumspection and listen to its proponents with even greater
scepticism than is usually necessary.  This is particularly true when
the gulf between observers and theoreticians is as wide as it usually
is here.  Either side may be more inclined to accept the claims of the
other than they should.  As an extra-galactic observer addressing a
mostly theoretical audience I want to emphasise the very many caveats
that should always be attached to the observational side of this
field.  I do so as a friend and admirer of George Ellis who has one of
the few minds capable of bridging the gulf.

\section{THE OBSERVATIONS WHICH BEAR ON COSMOLOGY}

The observations which bear on cosmology are, for such a grandiose
subject, extremely sparse.  I count only about a dozen which probably
bear - most of them stumbled upon by accident (see Table 1).  And they
are {\it observations} not controlled experiments which therefore
means that they cannot compare with the thousands of particle physics
experiments upon which the Standard Model is based.

\vspace{5mm}

{\bf Table 1}

ALL THE OBSERVATIONS WHICH BEAR ON COSMOLOGY

\vspace{5mm}

\begin{tabular}{ll}
1.&The dark sky background.*\\
2.&Isotropy of galaxy counts.\\
3.&Magnitude-Redshift diagram for galaxies.*\\
4.&Approx equivalence between 1/H$_0$ and $\tau_{\rm stars}$,
  $\tau_{\rm elements}$.*\\
5.&Existence of CBR.*\\
6.&Isotropy of CBR.*\\
7.&BB spectrum of CBR.\\
8.&Measured fluctuations in CBR?\\
9.&Abundance of Helium.*\\
10.&Abundance of Deuterium.*\\
11.&Magnitude-redshift diagram for supernovae.\\
12.&Existence of walls and voids in LSS.*\\
13.&Radio source-counts.*?
\end{tabular}

\vspace{5mm}

*Serendipitous.  ? = of questionable relevance.

\vspace{5mm}

\section{THE SPECIFIC DIFFICULTIES OF COSMOLOGY}\

Table 2 lists some of the special difficulties which cosmology has to
face as a science.  They are mostly obvious but it is worth
emphasising one or two:

\vspace{5mm}

{\bf Table 2}

PARTICULAR DIFFICULTIES FOR COSMOLOGY AS A SCIENCE

\vspace{5mm}

\begin{tabular}{ll}
1.&Only one Universe.\\
2.&Universe opaque for 56/60 decades since Planck era.\\
3.&Need to extrapolate physics over huge distances.\\
4.&Need to work with what we can currently detect. [But \ldots]\\
5.&Local background very bright.\\
6.&Distances very hard to determine (standard candles).\\
7.&Observational Selection insidious.\\
8.&Distant galaxies hard to measure and interpret unambiguously.\\
9.&Luminosity Functions unreliable.\\
10.&Geometry, astrophysics and evolution often entangled.\\
11.&Physics of early Universe unknown (and unknowable?)\\
12.&Human time-frame so short compared to cosmic.\\
13.&Origin of inertia.\\
14.&The singularity.
\end{tabular}

\vspace{5mm}

\renewcommand{\theenumi}{\Alph{enumi}}
\renewcommand{\labelenumi}{(\theenumi)}

\begin{enumerate}
\item There is only one Universe!  At a stroke this removes from our
  armoury all the statistical tools that have proved indispensable for
  understanding most of astronomy.
\item The Universe has been opaque to electromagnetic radiation for
  all but 4 of the 60 decades of time which stretch between the Plank
  era ($10^{-43}$ sec) and today ($10^{17}$ sec).  Since as much
  interesting physics could have occurred in each logarithmic decade,
  it seems foolhardy to claim that we will ever know much about the
  origin of the cosmos, which is lost too far back in the logarithmic
  mists of Time.  Even the Large Hadron Collider will probe the
  microphysics back only as far as $10^{-10}$ secs). [2].
\item Cosmology requires us to extrapolate what physics we know over
  huge ranges in space and time, where such extrapolations have
  rarely, if ever, worked in physics before.  Take gravitation for
  instance..  When we extrapolate the Inverse Square Law. ( - dress it
  up how you will as G.R.) from the solar system where it was
  established, out to galaxies and clusters of galaxies, it simply
  never works.  We cover up this scandal by professing to believe in
  ``Dark Matter'' - for which as much independent evidence exists as
  for the Emperor's New Clothes.
\item Objects at cosmologically interesting distance are exceedingly
  faint, small and heavily affected by factors such as
  redshift-dimming and {\it k}-corrections, so it will obviously be
  very difficult, if not impossible, to extract clear information
  about geometry, or evolution, or astrophysics - all of which are
  tangled up together.
\item Observational astronomy is all about the {\it contrast} between
  an object and its background [3] - both the background of the local
  Universe and the background noise in our instruments, which are
  never perfect.  Almost all the galaxies we know of are just
  marginally brighter than the terrestrial sky - either extraordinary
  good fortune, or more likely a signal that far more are hidden
  beneath it [4,5,6].  In other words we are in this, as in all other
  facets of observational astronomy, hapless victims of
  ``Observational Selection'' - an area in which George Ellis has done
  some brilliant work [7].  The sky isn't dark.  Even at the darkest
  site of Earth the unaided eye can pick up 50,000 photons a second
  coming from an area of ``dark sky'' no larger than the full moon.
  Bigger telescopes are all very well - but they pick up more unwanted
  foreground light, as well as background signal.  When you think that
  the galaxies at a redshift {\it z} of 2 should be dimmer by $(1 +
  z)^4 \sim 100$, and by another large but uncertain factor for the
  {\it k}-correction [i.e. band-pass shifting], it is more than a
  wonder to me that we can see anything of them at all.  Ordinary
  galaxies at that redshift should be hundreds of times dimmer per
  unit area than our sky!  It is also sobering to realise that only
  one per cent of the light in the night sky comes from beyond our
  Galaxy.
\item The tragedy of astronomy is that most information lies in
  spectra, and yet you need to collect between 100 and 1000 times more
  radiation to get a spectrum than to see an image.  Thus most of the
  faint galaxies which may have cosmological stories to tell must
  remain, in spectroscopic terms, tantalisingly out of earshot.  If
  history is anything to go by little good will come of the thousands
  of nights of big-telescope time now being lavished on the intriguing
  objects first seen with the Space Telescope, and made famous through
  the Hubble Deep Field.  We will probably learn more cosmology from
  studying the surprising and diverse histories of star-formation that
  Hubble is finding among galaxies in the Local Group [8].

In summary we have very few observations, most of them were accidently
made, and all are subject to observational selection.  It is therefore
outrageous to claim a comparison with all the carefully controlled
experiments made by particle physicists.  And even if we do get a
perfect map of the Cosmic Background Radiation it will only be a map
of a moment in time.  Celestial mechanics is very precise - but it
doesn't tell us how the solar system was formed.

\end{enumerate}

\section{THEORY AND OBSERVATIONS}

Martin Harwit [9] has argued that we cannot have made more than ten
per cent of the crucial discoveries in Astronomy.  He uses what John
Barrow aptly calls `the proof-readers argument'.  If two independent
readers look at a manuscript then it is possible to estimate, by
comparing their different results, how many errors there must be in
total, including those not identified.  In an analogous way two
independent astronomical channels (say optical and X-ray) can be used to
examine the Universe and a comparison of their separate key
discoveries will yield an estimate of the numbers still to be found.

In any case with so little data to work on it shouldn't be too
difficult to devise a plausible theory to account for them.  It is,
however, sobering to compare the cosmological situation with the
history of other sciences.

Take geology.  Men were living on the earth for millions of years, and
quarrying rock, digging mines and canals and puzzling over its fossils
for thousands of years, before unexpected palaeomagnetic patterns
revealed for certain the key idea of Continental Drift.

In stellar physics two thousand years elapsed between Hipparcos's
speculations and Bessel's first measurement of a stellar distance.
Seventy years later the statistical patterns in the H-R diagram  led
to our understanding of stellar structure.

However the closest comparison comes from my own field of galaxy
astronomy which is, as an observational science, almost exactly
contemporary with cosmology.  Although we now have good spectra and
images of thousands of galaxies the list of fundamental things we
don't know about them (Table 3) is far more striking that the list of
things we do.

\vspace{5mm}

{\bf Table 3}

WHAT WE DON'T KNOW ABOUT GALAXIES

\vspace{5mm}

\begin{tabular}{ll}
1.&How our knowledge is warped by Selection Effects.\\
2.&What they are mostly made of. (Dark Matter?)\\
3.&How they formed - and when.\\
4.&How much internal extinction they suffer from.\\
5.&What controls their global star-formation rates.\\
6.&What parts their nuclei and halos play.\\
7.&If there are genuine correlations among their global properties.\\
8.&How they keep their gas/star balances.
\end{tabular}

\vspace{5mm}

Of course these are only arguments by analogy.  The optimistic
cosmologist can always counter argue [I don't know how] that the
Universe in the large is a great deal simpler than its constituent
parts.

\section{THE COSMOLOGIST'S CREDO}

The cosmologist, who would also be a scientist, must surely subscribe
to at least the following assumptions:

\begin{enumerate}
\item ``Speculations are not made which cannot, at least in principle,
  be compared with observational or experimental data, for tests''
  [the NON-THEOLOGICAL assumption].
\item ``The portion of the Universe susceptible to observation is
  representative of the cosmos as a whole''. [The `GOOD LUCK'
  assumption].
\item ``The Universe was constructed using a significantly lower
  number of free parameters than the number of clean and independent
  observations we can make of it''. [The `SIMPLICITY' assumption].
\item ``The Laws of Physics which have significantly controlled the
  Universe since the beginning are, or can be, known to us from
  considerations {\it outside cosmology itself} i.e. we can somehow
  know the laws which operated during the 56/60 electromagnetically
  opaque decades''.  [The `NON-CIRCULARITY' assumption].

Finally the really wishful cosmologist who believes the final answers
are just around the corner must confess to the following extra creed:
\item ``We live in the first human epoch which possesses the
  technical means to tease out the crucial observations''.  (As
  opposed to Hipparcos and parallax, Helmholz and the age of the Earth,
  Wegener and palaeomagnetic drift) [The `FORTUNATE EPOCH'
  assumption.]

\end{enumerate}

I can see very little evidence to support any of the last 4
assumptions while it is dismaying to find that some cosmologists, who
would like to think of themselves as scientific, are quite willing to
abrogate the first.

\section{THE PATHOLOGIES OF COSMOLOGY}

\begin{enumerate}
\item  Cosmology must be the slowest moving branch of science.  The
  number of practitioners per relevant observation is ridiculous.
  Consequently the same old things have to be said by the same old
  people (and by new ones) over and over and over again.  For instance
  ``Cold Dark Matter'' now sounds to me like a religious liturgy which
  its adherents chant like a mantra in the mindless hope that it will
  spring into existence.  Much of cosmology is unhealthily self-referencing 
  and it seems to an outsider like myself that cosmological fashions and
  reputations are made more by acclamation than by genuine scientific
  debate.
\item There is a serious problem with the cost of astronomical
  spacecraft.  An instrument capable of cosmologically interesting
  observations may cost half a billion dollars or more.  There is
  therefore an insidious temptation to overclaim what they will see
  [1].  This, however, is a dangerous game which can blow up in your
  face, as proponents of the Supercollider were to find out.
\item There is something beguiling and yet fallacious about working
  on ``the faintest objects ever observed'' even though, by
  definition, they contain ``the least information ever detected''.
  During my working life a major fraction of the prime time on all
  large telescopes has been devoted to the study of objects right at
  the horizon, with, or so it seems to me, very little result.  To be
  rude about it, statistical studies of faint objects can keep a
  career going for ages without the need for a single original thought
  - or indeed a genuinely clear result.  The jam is always just around
  the next corner.
\item As particle physics has become paralyzed by its escalating cost
  many particle theorists have `moved over' into cosmology, wishfully
  thinking of the Universe as `The great Accelerator in the Sky'.
  Alas they are mostly not equipped with the astronomical background
  to appreciate how `soft' an observational, as opposed to an
  experimental science, has to be.  But they have only to look at the
  history of astronomy and at some of the howlers we have made (Table
  4) to find out.

\vspace{5mm}

{\bf Table 4}

SOME HISTORICAL MISTAKES IN COSMOLOGY

\vspace{5mm}

\begin{tabular}{ll}
1.&`Early' cosmologies - e.g. Genesis, Hindu, \ldots\\
2.&Many unsound explanations for dark sky (up to 1960).\\
3.&Assumption of a static Universe.\\
4.&Original expansion claim based on unsound statistics (Hubble).\\
5.&H$_0$ wrong by factor $\sim 10$ for 25 years.\\
6.&Universe measured to be younger than stars.\\
7.&CBR not recognised for 25 years [McKellar 1942, Gamov\ldots\\
8.&Radio-source counts misinterpreted due to use of fallacious
statistics.\\
9.&Mass of neutrinos forgotten/ignored for 40 years.\\
10.&Sandage's ``search for 2 numbers'' forgot evolution.\\
11.&Horizon/flatness problems virtually ignored before a possible
solution appeared.
\end{tabular}

\vspace{5mm}

\item Despite our intuitions very many Inverse Problems (and astronomy
  is very largely an Inverse Problem) are not well posed. [10].  For
  example when the HST was found to be spherically aberrated half the
  astronomical community claimed that the images could be restored by
  mathematical `deconvolution'.  But they could not be - because the
  problem is ill posed; the highest resolution information will be
  swamped by the highest frequency noise during the inversion - it is
  a fundamental property of numerical differentiation.  Only very
  high signal-to-noise data (a luxury astronomers rarely enjoy) can be
  deconvolved successfully.  Likewise, I suspect that the
  multiparticle simulations beloved of certain numerical cosmologists
  are extremely ill-posed.  They start off with a whole lot of CDM
  `dots', the dots apparently form filaments under the force of
  gravity - as they are bound to do according to Zeldovich's simple
  back-of-the-envelope analysis, and we are supposed to admire the
  result.  What result?  That to me is the question.  Presumably we
  are supposed to compare the dots with real structures and infer some
  properties of the physical Universe.  In my opinion it is nothing
  more than a seductive but futile computer game.  What about the
  gas-dynamics, the initial conditions, the star-formation physics,
  evolution, dust, biasing, a proper correlation statistic, the
  feedback between radiation and matter \ldots ?  Without a good stab
  at all these effects `dotty cosmology' is no more relevant to real
  cosmology than the computer game `Life' is to evolutionary biology.
\item However, the most unhealthy aspect of cosmology is its unspoken
  parallel with religion.  Both deal with big but probably unanswerable
  questions.  The rapt audience, the media exposure, the big
  book-sale, tempt priests and rogues, as well as the gullible, like
  no other subject in science.  For that reason alone other scientists
  simply must treat the pretensions of cosmology, and of professional
  cosmologists, with heightened scepticism, as I am attempting to do
  here.
\end{enumerate}

\section{COSMOLOGY IN PERSPECTIVE}

Of course we would all love to know of the fate of the Universe, just
as we'd love to know if God exists.  If we expect science to provide
the answers though, we may have to be very patient - and literally wait
for eternity.  Alas professional cosmologists cannot afford to wait
that long.  For that reason the word `cosmologist' should be expunged
from the scientific dictionary and returned to the priesthood where it
properly belongs.

I'm not suggesting that cosmology itself should be abandoned.  Mostly
by accident it has made some fascinating, if faltering progress over
the centuries.  And if we are patient and build our instruments to
explore the Universe in all the crevices of parameter space, new clues
will surely come to hand, as they have in the past, largely by
accident.  But we should not spend too many of our astronomical
resources in trying to answer grandiose questions which may, in all
probability, be unanswerable.  For instance we must not build the Next
Generation Space Telescope as if it was solely a cosmological
machine.  We should only do that if we are confident of converging on
``the truth''.  If we build it to look through many windows we may yet
find the surprising clues which lead us off on a new path along the
way.

Above all we must not overclaim for this fascinating subject which, it
can be argued, is not a proper science at all.  Rutherford for
instance said ``Don't let me hear anyone use the word `Universe' in my
department''.  Shouldn't we scientists be saying something like this
to the general public:

\vspace{5mm}

\noindent{\it ``It is not likely that we primates gazing through bits
  of glass for a century or two will dissemble the architecture and
  history of infinity.  But if we don't try we won't get anywhere.
  Therefore we professionals do the best we can to fit the odd clues
  we have into some kind of plausible story.  That is how science
  works, and that is the spirit in which our cosmological speculations
  should be treated.  Don't be impressed by our complex machines or
  our arcane mathematics.  They have been used to build plausible
  cosmic stories before - which we had to discard afterwards in the
  face of improving evidence.  The likelihood must be that such
  revisions will have to occur again and again and again.''}

\vspace{5mm}

I apologise for such a highly opinionated attack, but it does appear
to me that the pendulum has swung much too far the other way.  Surely
the `burden of proof' ought to rest squarely on the proponents of what
will always be a fascinating but suspect subject.

\section*{REFERENCES}

\renewcommand{\theenumi}{\arabic{enumi}}
 \renewcommand{\labelenumi}{\theenumi.}

 \begin{enumerate}
 \item Hu, W., Sugiyama, N. And Silk, J., 1997, {\it Nature}, 386,
   37.
 \item Rees, M., 1995, {\it Perspectives in Astrophysics Cosmology},
   {\it CUP}, 109.
 \item Condon, J., 1998, IAU Symposium 179, (Kluwer), p19.
 \item Disney, M. J., 1976, {\it Nature}, 263, 573.
 \item Impey, C. And Bothun, G., 1998, Ann. Revs. Astron. Astrophys.
 \item Disney, M. J., 1998, {\it IAU Colloquium 171}, (Kluwer), p11.
 \item Ellis, F. G. R., Perry, J. J., And Sievers, A. W., 1984, AJ, 89,
   1124
 \item Mateo, M. G., 1998, Ann. Revs. Astron. Astrophys., p435
 \item Harwit, M., 1981, {\it Cosmic Discovery}, {\it Harvester Press
     UK}, p231.
 \item Craig and Brown, 1986, {\it Inverse Problems in Astronomy (Adam
     Hilger; Bristol)}.
 \end{enumerate}
\end{document}